\documentclass[%
 reprint,
superscriptaddress,
a4paper,
amsmath,amssymb,amsfonts,
floatfix,
]{revtex4-1}

\usepackage{graphicx}
\usepackage[colorlinks,
citecolor=blue,
]{hyperref}
\usepackage{comment}
\usepackage{xcolor}
\usepackage{wasysym}
\usepackage{color}

\newcommand{\lc}{\ell_c}

\newcommand{\upd}{\mathrm{d}}

\newcommand{\cosinv}{\cos^{\mathrm{-1}}}

\newcommand{\sininv}{\sin^{\mathrm{-1}}}

\newcommand{\cosech}{\mathrm{cosech}}
\newcommand{\be}{\begin{equation}}
\newcommand{\ee}{\end{equation}}
\newcommand{\sgn}{\mathrm{sgn}}

\definecolor{myGreen}{rgb}{0,0.5,0}

\begin{document}

\title{Maximal liquid bridges between horizontal cylinders}

\author{Himantha Cooray}
\affiliation{Institute of Theoretical Geophysics, Department of Applied Mathematics and Theoretical Physics, University of Cambridge, Centre for Mathematical Sciences, Wilberforce Road, Cambridge CB3 0WA, UK}
\email{himantha@cantab.net}
\homepage{http://himantha.freeshell.org/}
\author{Herbert E. Huppert}
\affiliation{Institute of Theoretical Geophysics, Department of Applied Mathematics and Theoretical Physics, University of Cambridge, Centre for Mathematical Sciences, Wilberforce Road, Cambridge CB3 0WA, UK}
\affiliation{Faculty of Science, University of Bristol, Bristol BS8 1UH, UK}
\affiliation{School of Mathematics and Statistics, University of New South Wales, Sydney, NSW 2052, Australia}
\author{Jerome A. Neufeld}
\affiliation{Institute of Theoretical Geophysics, Department of Applied Mathematics and Theoretical Physics, University of Cambridge, Centre for Mathematical Sciences, Wilberforce Road, Cambridge CB3 0WA, UK}
\affiliation{BP Institute, Bullard Laboratories, University of Cambridge, Madingley Road, Cambridge CB3 0EZ, UK}
\affiliation{Department of Earth Sciences, Bullard Laboratories, University of Cambridge, Madingley Road, Cambridge CB3 0EZ, UK}

\begin{abstract}
We investigate two--dimensional liquid bridges trapped between pairs of identical horizontal cylinders. The cylinders support forces due to surface tension and hydrostatic pressure which balance the weight of the liquid. The shape of the liquid bridge is determined by analytically solving the nonlinear Laplace--Young equation. Parameters that maximize the trapping capacity (defined as the cross--sectional area of the liquid bridge) are then determined. The results show that these parameters can be approximated with simple relationships when the radius of the cylinders is small compared to the capillary length. For such small cylinders, liquid bridges with the largest cross sectional area occur when the centre--to--centre distance between the cylinders is approximately twice the capillary length. The maximum trapping capacity for a pair of cylinders at a given separation is linearly related to the separation when it is small compared to the capillary length. The meniscus slope angle of the largest liquid bridge produced in this regime is also a linear function of the separation. We additionally derive approximate solutions for the profile of a liquid bridge making use of the linearized Laplace--Young equation. These solutions analytically verify the above relationships obtained for the maximization of the trapping capacity.
\end{abstract}

\maketitle

\section{Introduction}

The trapping of a fluid in contact with a solid is a general problem with applications in biological, engineering, industrial and geological processes. Generally, a volume of liquid trapped by two or more solid surfaces and immersed in a different fluid is called a ``liquid bridge''. The trapping is achieved by balancing the weight of the liquid with the surface tension forces acting along the three--phase contact lines and the forces of hydrostatic pressure exerted on the solid--liquid contact surfaces. A detailed review of liquid bridges can be found in \citet{Butt2009a}. Liquid bridges are a very common occurrence in granular matter and porous media. Examples include trapping of water in sand, which acts as an adhesive in sand castles \citep{Schiffer2005}, and capillary trapping of supercritical carbon dioxide in porous rocks \citep{Juanes2009} during carbon dioxide sequestration.

In this paper, we study two-dimensional liquid bridges produced between pairs of horizontal cylinders. A study in this simplified geometry is a first step in the detailed understanding of trapping in porous media. It can also give insights into the behaviour of a three--dimensional liquid bridge trapped between cylindrical rods. Liquid absorption to textiles \citep{Lukass2003} and retention of water droplets on spider webs are common examples of trapping in this geometry. Additionally, it has recently been proposed as a method of handling and mixing small volumes of liquid in analytical research \citep{Cheong2013177}. \citet{Princen1970} and \citet{Lukass2003} solved this problem in two dimensions neglecting the effects of gravity. Such solutions lose their accuracy as the amount of trapped liquid increases. Although three--dimensional profiles of trapped droplets have been studied experimentally \citep{Protiere2013,Duprat2012} and numerically \citep{Wu2010,Bedarkar2010}, there is no straightforward method to determine how much liquid a given system can trap. 

\begin{figure*}[t]
\centering
\includegraphics [width=8cm]{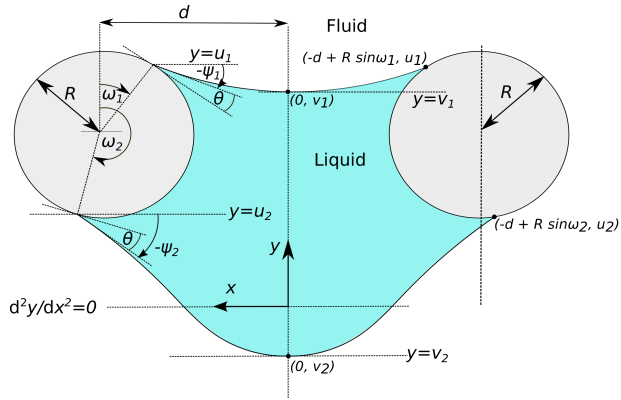}
\caption{A liquid bridge formed between a pair of horizontal cylinders. $\theta$ is the contact angle, $\omega_1$ and $\omega_2$ are the angles from the vertical where the liquid meets the cylinder and $\psi_1$ and $\psi_2$ are the interfacial slope angles, which are positive if measured counter-clockwise. All the lengths are nondimensionalized by dividing by the capillary length. The height $y$ is proportional to the pressure of the liquid at the liquid--fluid interface relative to the pressure of the fluid. At $y=0$, the pressure difference between the two phases and the interfacial curvature are $0$. $R$ is the radius of the cylinders and $d$ is the half distance between their centres. 
}
\label{figSchematic}
\end{figure*}

Capillary trapping in other related geometries has been studied using a variety of methods. \citet{Urso1999a} analysed trapping of a liquid in a two--dimensional porous medium comprised of horizontal cylinders. They studied trapping in the limit of small liquid volumes, where gravitational effects can be neglected and the liquid--fluid interfaces may be approximated by circular arcs. \citet{Chen1992} determined the shape of a three--dimensional liquid bridge trapped between vertical plates using a perturbation method in which the weight of the liquid was neglected, and calculated numerically, using a finite element method, cases in which the weight was incorporated. 
While a two--dimensional liquid bridge is approximately symmetric in the vertical if its weight is close to zero, the shape becomes significantly asymmetric when more liquid is added. 
The shape of the lower interface in this regime can be modelled as a pendant drop. Profiles of pendant drops have been studied extensively for two--dimensional \citep{Pitts1973,Majumdar1976} and axially symmetric \citep{Orr1975,Boucher1975} cases. 
Although the above solutions take all the physical parameters into account, they are either analytical solutions that give complicated expressions or numerical solutions and, as a result, do not provide direct expressions to determine the trapping capacity.

The study in this paper starts with an exact solution for the profile of a two--dimensional liquid bridge of arbitrary volume. Results obtained using this solution show very simple approximate relationships governing the maximum trapping capacity: the maximum trapping capacity is linearly related to the separation between the cylinders when the separation is small compared to the capillary length; and the separation that produces the largest trapping capacity is twice the capillary length. We then analytically verify these limiting relationships using several approximate solutions for the shape of a liquid bridge.

\section{Theoretical Setting}

We consider a two dimensional, horizontally  symmetric liquid bridge produced between a pair of identical horizontal cylinders as shown in figure \ref{figSchematic}. The weight of the liquid is balanced by the forces of surface tension and the reaction to the hydrostatic pressure exerted by the cylinders. Both liquid--fluid interfaces of the liquid bridge meet the cylinders at a fixed contact angle $\theta$, which is in practice locally determined by the fluid and solid surface energies. The interfacial slope angles at the contacts are given by $\psi_i$, where the subscript $i=1$ denotes the upper interface and $i=2$ denotes the lower interface, and $\psi_i$ is positive if the interface slopes upwards leaving the cylinder. The point of contact between a cylinder and an interface is denoted by the angle $\omega_i$ to the vertical. The following relationships between $\psi_i$, $\theta$ and $\omega_i$ are obtained by consideration of the geometry of the system
\begin{align}
\psi_1 &= \theta-\omega_1, \label{psi1}\\
\psi_2 &= \pi- \theta-\omega_2. \label{psi2}
\end{align}
The shape of each liquid interface of the liquid bridge is governed by the nonlinear Laplace--Young equation which relates the pressure difference across the interface to its curvature. If the height $Y$ of the interface is given as a function of the horizontal position $X$ by $Y=G(X)$, the Laplace--Young equation is written as 

\begin{equation}
  \label{eq:LYEx}
  Y = \lc^2~P(X)\frac{G_{XX}}{\left[G_X^{~2}+1\right]^{3/2}},
\end{equation}
where the subscripts denote derivatives and the capillary length defined as 
\begin{equation}
  \label{eq:LcDefine}
  \lc \equiv \sqrt{\frac{\gamma}{\Delta \rho~ g}},
\end{equation}
in which $\gamma$ is the liquid--fluid interfacial tension, $\Delta \rho$ is the density difference between the liquid and the fluid and $g$ is the acceleration due to gravity. 
$P(X)=\pm 1$ depending on whether the liquid phase is below or above the fluid phase, and it is defined as
\begin{equation}
  \label{eq:SX}
  P(X) = \sgn\left\{ 
D[X,~G(X)-\delta]-D[X,~G(X)+\delta] 
\right\},
\end{equation}
where $D(X,Y)$ is the density at a location $(X,Y)$ covered by a fluid, which is assumed to be constant within each phase, and $\delta$ is a positive infinitesimal length.

\begin{figure*}[t!]
\centering
\includegraphics [width=7cm] {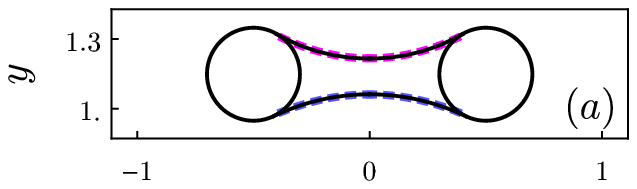}\\ 
\vspace{3 mm}
\includegraphics [width=7cm] {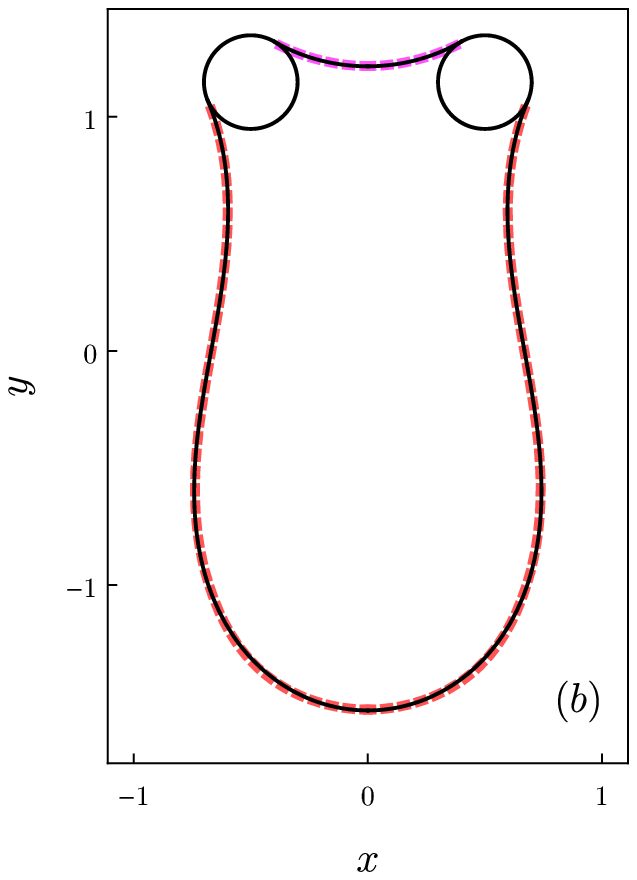} 
\caption{Shapes of two liquid bridges obtained using exact and approximate solutions of the Laplace--Young equation. Some values of $\psi_1$ can produce two different liquid bridges because \eqref{bound3All} can have two solutions for $\psi_2$. Both liquid bridges shown here are obtained using the same input parameters $R=0.2, d=0.5, \theta= 0$ and $\psi_1=-\pi/6$ and represent two solutions for $\psi_2$, i.e. $0.44$ in $(a)$ and $-1.17$ in $(b)$. The black solid curves (--) are obtained from the solution to the nonlinear Laplace--Young equation \eqref{eqfgboth}. 
The magenta dashed curves (\textcolor{magenta}{\bf{- -}}) show an approximation \eqref{eqUpperLin} for the shapes of the upper interfaces of the liquid bridges obtained by solving the linearized Laplace--Young equation. The blue dashed curve (\textcolor{blue}{\bf{- -}}) in (a) shows a similar approximation \eqref{eqLowerLin} for the lower interfaces which is valid when the interfacial slopes are small. The red dashed curve (\textcolor{red}{\bf{- -}}) in (b) is a composite approximation for the shape of the lower interface, valid for distended liquid bridges, given by \eqref{compositeSol}. The results show very good agreement between the exact and approximate solutions.
}
\label{figTwoSolutions}
\end{figure*}

Due to the symmetry of the system, we only need to solve for a half of the bridge to determine its full shape. (In the solution presented here, we only consider the left side). However, writing the Laplace--Young equation in the form of  (\ref{eq:LYEx}) has several drawbacks. First, it cannot be solved by direct integration and, secondly, the shape of the lower interface can be multivalued relative to $X$ and also $P(X)$ can change sign within a single fluid interface (for example, consider the lower fluid interface of the liquid bridge shown in figure \ref{figTwoSolutions}(b)). These problems can be avoided by instead expressing the interfacial shape as a function of $Y$. It is also convenient to nondimensionalize all the lengths with respect to the capillary length and define $x=X/\lc$ and $y=Y/\lc$. The interfacial shape can then be written as
\begin{equation}
x  = f(y),
\label{eq:xISfy}
\end{equation}
where  $x=0$ is the axis of symmetry and  $y=0$ represents the vertical coordinate at which $\upd^2 f / \upd y^2=0$, which is not known a priori and has to be determined as a part of the solution. The nondimensionalized Laplace--Young equation is 
\begin{equation}
y  = p(y) \frac{f_{yy}}{\left(f_y^2+1\right)^{3/2}},
\label{LYE}
\end{equation}
with $p(y)=\pm 1$ according to the relative positions of the liquid and fluid. Since the interfacial shape is defined as a function of the vertical coordinate, $p(y)$ is now determined by whether the liquid phase is located in the right hand side or left hand side of the fluid phase, so that
\begin{equation}
  \label{eq:pDefFull}
  p(y) = \sgn\left\{
D_n[f(y)-\epsilon,~y]-D_n[f(y)+\epsilon,~y]
   \right\}  
\end{equation}
as $\epsilon \rightarrow 0$, from above, where $D_n(x,y)$ is the fluid density at a location $(x,y)$ which is specified in terms of the nondimensionalized coordinates. Since only a half of a liquid bridge is to be solved, $p$ is constant within each interfacial segment we consider and it depends only on the direction of the meniscus slope at the contact point
\be
p=
\begin{cases}
\sgn(\psi_1) &\mbox{for the upper meniscus}\\ 
-\sgn(\psi_2) &\mbox{for the lower meniscus}.\\ 
\end{cases}
\label{eqp}
\ee

The liquid bridge shown in figure \ref{figSchematic} is  trapped between cylinders of (nondimensionalized) radius $R$ and a centre-to-centre distance $d$. If the vertical coordinates of the contact point and middle point of each interface of the liquid bridge are $y=u_i$ and $y=v_i$ respectively, the interfacial slope angle defines a boundary condition at each contact point
\be
f_y(u_i) = -\cot(\psi_i),
\label{eqfyBC}
\ee
and the requirement for symmetry provides a boundary condition at the centre line
\begin{equation}
\lim_{y \to v_i} f_y =  -\sgn(\psi_i) ~\infty.
\label{eqfyMid}
\end{equation}
Finally, we impose that the free surfaces intersect the cylinder at the points
\begin{equation}
f(u_i)=d-R \sin \omega_i,
\label{bound3All}
\end{equation}
and are continuous across the centre line
\begin{equation}
f(v_i)=0.
\label{boundfvi}
\end{equation}

In the following section, we obtain a solution for the full shape of the liquid bridge given $R$, $\theta$, $d$ and $\omega_1$ (or $\psi_1$) and predict $\omega_2$, $u_i$ and $v_i$ as part of the solution.

\section{Exact Solution of the nonlinear Laplace--Young equation}

\label{sec:exactsolution}
The Laplace--Young equation given in \eqref{LYE} may be integrated and rearranged to obtain
\begin{equation}
f_y= p' \frac{\frac{1}{2}y^2- a_i}{\sqrt{1-\left(\frac{1}{2}y^2- a_i\right)^2}},
\label{dfdy}
\end{equation}
where $a_i$ is a constant of the integration and $p'=\pm 1$. To determine the value of $p'$, we differentiate the above equation to obtain
\begin{equation}
f_{yy}= p' \frac{y}{\left[1-\left(\frac{1}{2}y^2- a_i\right)^2 \right]^{\frac{3}{2}}}.
\label{dfdyy}
\end{equation}
Comparison of this result with ~(\ref{LYE}) shows that
\be
p'=p.
\ee

Substitution of $f_y$ given by \eqref{dfdy} into \eqref{eqfyMid}, which denotes the meniscus slope at the mid--point of each interface, yields

\begin{equation}
a_i = \frac{1}{2}v_i^2 +q,
\label{eqa1}
\end{equation}
where 
\be
q=p ~\sgn(\psi_i).
\label{eqq1}
\ee
The value of $p$ in \eqref{eqp} is combined with \eqref{eqq1} to produce
\be
q=
\begin{cases}
1 &\mbox{for the upper meniscus}\\ 
-1 &\mbox{for the lower meniscus}.\\ 
\end{cases}
\label{eqq}
\ee

We then combine \eqref{eqfyBC}, which gives the meniscus slope at a contact point, with \eqref{dfdy} and \eqref{eqa1} to obtain
\begin{equation}
u_i^2 =  v_i ^2 + 2 q \left(1-\cos \psi_i \right).  
\label{equivi}
\end{equation}
The general shape of an interface is determined by integration of  \eqref{dfdy}. This integration is carried out using the substitution
\begin{equation}
\frac{1}{2}y^2-a_i = \cos \alpha,
\label{ysubs}
\end{equation}
which transforms \eqref{dfdy} to
\begin{equation}
f_\alpha = -p \frac{cos \alpha}{2\sqrt{a_i+\cos\alpha}}.
\end{equation}
The interface may therefore be described completely by the expression
\begin{align}
f(y) = p \Big\{ &-\sgn(y)~g(y)+ \left[ \sgn(y)-\sgn(v_i) \right]~g(0)  \nonumber\\
& +\sgn(v_i)~g(v_i) \Big\},
\label{eqfgboth}
\end{align}
where
\vspace{5 mm}
\begin{widetext}
\begin{align}
\label{eqJBoth}
g(y) =  &\sqrt{2(1+q)+v_i^2} ~E \left[ \frac{1}{2} \cosinv \left( \frac{y^2-v_i^2}{2} -q \right), \frac{4}{2(1+q)+v_i^2} \right]  \nonumber\\
&-\frac{2q+v_i^2}{\sqrt{2(1+q)+v_i^2}}~ F \left[ \frac{1}{2} \cosinv \left( \frac{y^2-v_i^2}{2}-q \right), \frac{4}{2(1+q)+v_i^2} \right],
\end{align}
\end{widetext}
\noindent is given in terms of incomplete elliptic integrals
$E(\sigma,k)$ and $F(\sigma,k)$ \citep{Byrd1971}. 
This equation satisfies the boundary condition $f(v_i)=0$ and remains continuous at $y=0$.

According to the Laplace--Young equation, the pressure in the liquid side of the interface is higher than the pressure in the fluid side when a liquid surface is convex. As a result, a convex liquid surface corresponds to a negative $y$ and a concave liquid surface corresponds to a positive $y$. If the lower interface of the liquid bridge slopes downwards at the contact point (i.e. $\psi_2 < 0$), it has to  be convex at the mid point ($x=0$) to satisfy the symmetry. This makes $v_2$ negative. If $\psi_2$ is positive, the interface is concave in the middle and $v_2$ is therefore positive. Using a similar argument for the upper interface as well, one can obtain the following general relationship for a liquid bridge. 
\be
\sgn(v_i) = -q~ \sgn(\psi_i),
\label{sgnvi}
\ee
This equation can be used to eliminate ~$\sgn(v_i)$ from \eqref{eqfgboth} to obtain
\begin{align}
f(y) =p \Big\{ &-\sgn(y)~g(y)+ \left[ \sgn(y)+ q~ \sgn(\psi_i) \right]~g(0) \nonumber\\
 &-q~\sgn(\psi_i)~g(v_i) \Big\},
\label{eqfgbothpsi}
\end{align}
and $v_i$ can be eliminated from \eqref{eqJBoth} using \eqref{equivi} to produce
\begin{widetext}
\begin{align}
\label{eqJBothU}
g(y) =&\sqrt{2(1+q\cos \psi_i)+u_i^2} ~E \left[ \frac{1}{2} \cosinv \left( \frac{y^2-u_i^2}{2} -q \cos \psi_i\right), \frac{4}{2(1+q\cos \psi_i)+u_i^2} \right]  \nonumber \\
&-\frac{ 2q \cos \psi_1 + u_i^2 }{ \sqrt{2(1+q\cos \psi_i)+u_i^2} }~ F\left[ \frac{1}{2} \cosinv \left( \frac{y^2-u_i^2}{2} -q \cos \psi_i\right), \frac{4}{2(1+q\cos \psi_i)+u_i^2} \right].
\end{align}
\end{widetext}

We now use the boundary condition that defines the horizontal position of the contact point of each menisci given by \eqref{bound3All} to obtain a relationship between $\psi_i$ and $u_i$. 

The geometry of the cylinder gives the relationship between the vertical positions of the upper and lower contact points of the menisci
\begin{equation}
u_2 = u_1 - R \left( \cos \omega_1  -\cos \omega_2 \right),
\label{u1u2}
\end{equation}
from which $\omega_i$ can be replaced using \eqref{psi1} and \eqref{psi2} to obtain
\begin{equation}
u_2 = u_1 - R \left[ \cos \left(\theta-\psi_1\right) + \cos \left(\theta+\psi_2 \right) \right].
\label{u1u2psi}
\end{equation}
Equations \eqref{u1u2} and \eqref{bound3All} with $i=1$ and $2$ then represent three equations for $\psi_1$, $\psi_2$, $u_1$ and $u_2$. If any one of these four parameters is known, the other three can  be determined and the shapes of both the menisci can be found.

The following steps show the method used to determine the shapes of the liquid bridges in this paper.
 
\begin{enumerate}
\item  Select the upper point of contact with the cylinder $\omega_1$ and determine  $\psi_1$ using \eqref{psi1}, or select $\psi_1$ directly.
\item  Substitute  (\ref{sgnvi}), (\ref{eqfgboth}) and (\ref{eqJBothU}) into  \eqref{bound3All} and solve for $u_1$.
\item  Express $u_2$ as a function of $\psi_2$ using \eqref{u1u2psi}.
\item  Determine $\psi_2$ by solving \eqref{bound3All}, into which (\ref{sgnvi}),(\ref{eqfgboth}) and (\ref{eqJBothU}) are substituted.
\item  Determine $\omega_2$ using \eqref{psi2}.
\item  Obtain the shapes of the menisci using \eqref{eqfgboth}.
\end{enumerate}

For a given value of $\psi_1$, \eqref{bound3All} gives only one solution for $u_1$. However, for some values of $u_2$, the solution is multivalued, and thus can give two solutions for $\psi_2$ resulting in two different liquid bridges as shown in figure \ref{figTwoSolutions}. The first solution produces a liquid bridge with approximate vertical symmetry and the second solution produces a larger liquid bridge in which the lower interface is significantly distended, and as a result, contains a larger amount of liquid compared to the first. Both these solutions are equally valid.

\section{Approximate solutions for the shapes of the liquid interfaces}
\label{sectionApprox}

\begin{figure*}[p]
\centering
\includegraphics [height=5.9cm] {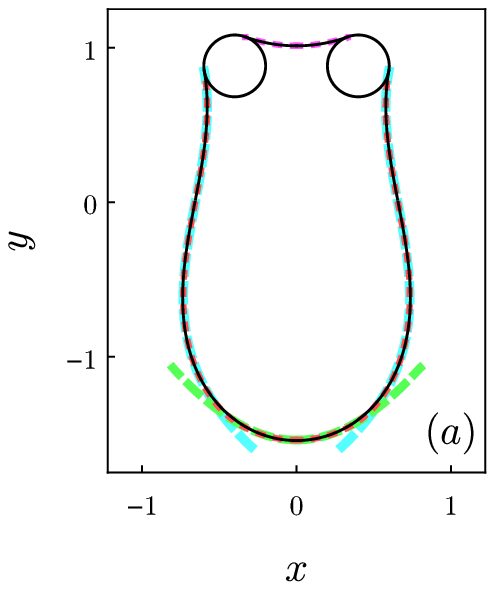}~~~
\includegraphics [height=5.9cm] {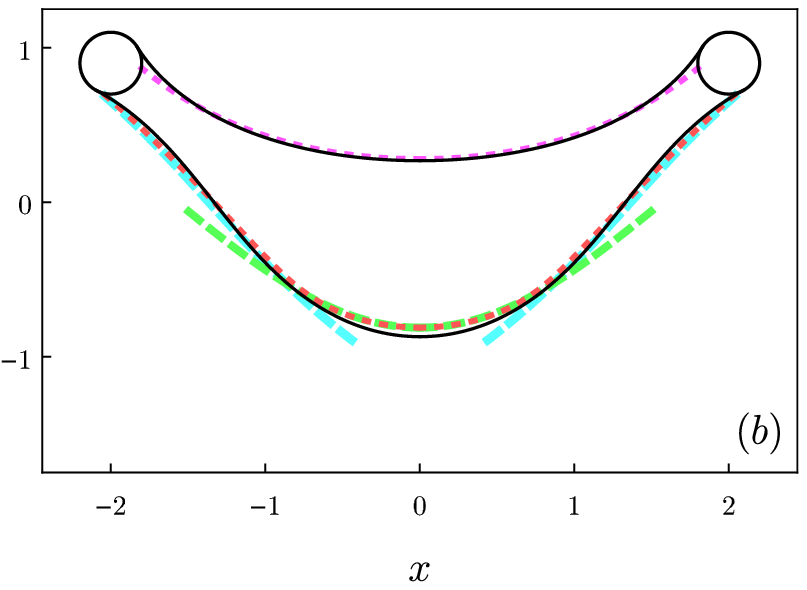}\\
\vspace{8 mm}
\includegraphics [height=8cm] {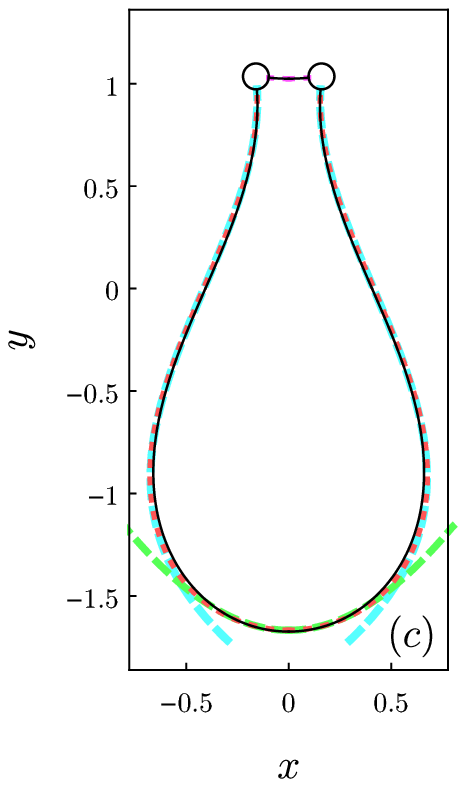}~~~~
\includegraphics [height=8cm] {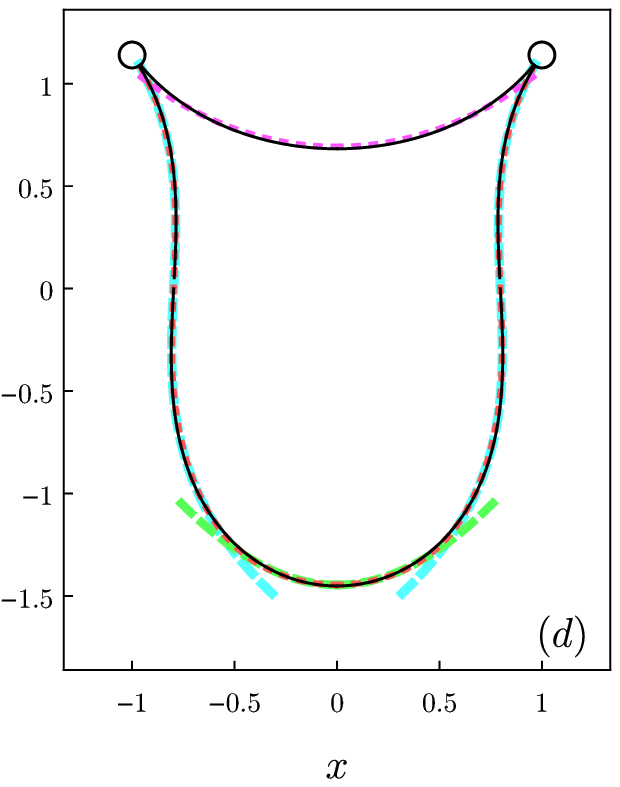}
\caption{$(a)$ and $(b)$ show two liquid bridges carrying the same amount of liquid (cross sectional area, $A=3.0$) between a pair of horizontal cylinders with $R=0.2$ and $\theta=0$ located at two different separations. The separation $d$ in $(a)$ is $0.4$, which gives $\psi_1=-\pi/9$ and $\psi_2=-1.40$. Parameters in $(b)$ are $d=2.0$, $\psi_1=-1.01$ and $\psi_2=-0.48$. 
$(c)$ and $(d)$ show shapes of liquid bridges corresponding to the maximum trapping capacities for a pair of cylinders with $R=0.063$ and $\theta=\pi/2$ at two different separations. The parameters are $d=0.16$, $A_\mathrm{max}=2.31$, $\psi_{1,A_\mathrm{max}}=-0.10$ and $\psi_2=-1.50$ in $(c)$ and $d=1.0$, $A_\mathrm{max}=3.24$, $\psi_{1,A_\mathrm{max}}=-0.88$ and $\psi_2=-1.00$ in $(d)$. 
The figures show results obtained using both exact and approximate solutions to the Laplace--Young equation. 
The black solid curves (--) are the solutions to the nonlinear Laplace--Young equation given by \eqref{eqfgbothpsi}. 
The magenta dashed curve  (\textcolor{magenta}{\bf{- -}}) is the approximation for the shape of the upper interface \eqref{eqUpperLin} obtained by solving the linearized Laplace--Young equation. The cyan dashed curve (\textcolor{cyan}{\bf{- -}}) and the green dashed curve (\textcolor[rgb]{0,1,0}{\bf{- -}})
are the approximations for the shapes of the upper part and the lower part of the lower interface given by \eqref{eqh} and \eqref{lowerMeniscusLowerPart} respectively.
The red dashed curve  (\textcolor{red}{\bf{- -}}) is the composite approximation for the shape of the lower interface \eqref{compositeSol} obtained by combining \eqref{eqh} and \eqref{lowerMeniscusLowerPart}.
There is excellent agreement between the  approximate and exact solutions when $\psi_1 \rightarrow 0$ and $\left| \psi_2 \right| \rightarrow \pi/2$. The composite approximation \eqref{compositeSol} covers both \eqref{eqh} and \eqref{lowerMeniscusLowerPart} very well.
}
\label{figShapes1}
\end{figure*}

\subsection{Shape of the upper interface as $|\psi_1| \to 0$}
\label{solutionSmallPsi1}

Expressing the shape of the upper meniscus by the function $y=j(x)$ and assuming the interfacial slopes to be small ($j_x \ll 1$), we can write the linearized Laplace--Young equation as
\begin{equation}
j = j_{xx}.
\end{equation}
Solution of this equation with the boundary condition $j_x(0)=0$ gives
\begin{equation}
j(x) = c_0 \cosh x,
\end{equation}
where $c_0$ is a constant to be determined. 
Since the vertical component of the surface tension force exerted by the cylinders at the contact points is equal to the weight of a liquid meniscus with vertical edges \citep{Vella2005, Keller1998}, the force balance may be written as
\begin{equation}
\int_0^{d-R \sin \omega_1}j ~\upd x = -\sin \psi_1.
\end{equation}
This gives the correct value for $c_0$, and so 
\begin{equation}
j(x) = -\sin \psi_1 \frac{\cosh x}{\sinh \left[d-R~ \sin (\theta-\psi_1)\right]},
\label{eqUpperLin}
\end{equation}
which is valid in the region where the meniscus slopes are small. If the absolute value of the meniscus slope angle $|\psi_1|$ is small, this solution is valid throughout the meniscus, and if $|\psi_1|$ is large, the solution is valid far (compared to $\lc$) away from the contact points. As a result, the approximation for $v_1$ obtained using \eqref{eqUpperLin} is in general more accurate than the approximation for $u_1$ obtained using the same equation. The height of the mid--point of the meniscus is therefore obtained using \eqref{eqUpperLin} as
\be
v_1 = -\sin \psi_1 ~\cosech \left[d-R~ \sin (\theta-\psi_1)\right],
\label{eqLinv1}
\ee
and $u_1$ is to be determined using \eqref{equivi}, which is a relationship between $u_1^2$ and $v_1^2$ derived from the nonlinear Laplace--Young equation. 
The upper interface cannot pass through $y=0$ because the interface is convex to the fluid side when $y<0$ and convex to the liquid side when $y>0$ according to the Laplace--Young equation. Therefore we have

\be
\sgn(u_1)=\sgn(v_1),
\label{eq:sgnu1}
\ee
where  $\sgn(v_1)$ is given by \eqref{sgnvi}.
Using \eqref{eqLinv1}, \eqref{eq:sgnu1} and \eqref{eqq} on \eqref{equivi}, we obtain
\begin{align}
u_1 = -&\sgn(\psi_1) \nonumber\\
& \times \Big\{\sin^2 \psi_1 ~\cosech^2  \left[d-R~ \sin (\theta-\psi_1)\right] \nonumber\\
&~~~~~~+ 2 \left(1- \cos \psi_1 \right)\Big\}^{1/2}
\label{eqLinu1}
\end{align}
for the contact height of the meniscus.

\subsection{Shape of the lower interface}

\subsubsection{Solution for small liquid volumes, $|\psi_2| \to 0$}
\label{solutionSmallPsi2}

In the limit of small liquid volumes, the upper and lower interfaces are nearly symmetric.
If the shape of the lower meniscus is given by $y=k(x)$, the linearized Laplace--Young equation is 
\be
k=-k_{xx}.
\label{linLYELow1}
\ee
This is solved in a manner similar to the upper interface to obtain
\begin{equation}
k(x) = -\sin \psi_2 \frac{\cos x}{\sin \left[d-R~ \sin (\theta+\psi_2)\right]},
\label{eqLowerLin}
\end{equation}
which gives
\be
v_2 = -\sin \psi_2 ~\csc \left[d-R~ \sin (\theta+\psi_2)\right].
\label{eqLinv2}
\ee
For small values of $|\psi_2|$ we have 
\be
\sgn(u_2)=\sgn(v_2).
\ee
Substitution of the above two equations into \eqref{equivi} and \eqref{sgnvi} produces
\begin{align}
u_2 =& \sgn(\psi_2) \nonumber\\
&\times \Big\{\sin^2 \psi_2 ~\csc^2 \left[d-R~ \sin (\theta+\psi_2)\right] \nonumber\\
&~~~~~~- 2 \left(1- \cos \psi_2 \right)\Big\}^{1/2}.
\label{eqLinu2}
\end{align}

Equation \eqref{eqLinu1} gives the value of $u_1$ for a given $\psi_1$. This is substituted into \eqref{u1u2psi} to express $u_2$ as a function of $\psi_2$,
\begin{align}
\label{equ2fromu1}
u_2 = - &\sgn(\psi_1) \nonumber\\
& \times \Big\{\sin^2 \psi_1 ~\cosech^2 \left[d-R~ \sin (\theta-\psi_1)\right] \nonumber\\
&~~~~~~+ 2 \left(1- \cos \psi_1 \right) \Big\}^{1/2}\nonumber\\
& - R \left[\cos (\theta-\psi_1) + \cos (\theta+\psi_2) \right].
\end{align}
Equations \eqref{equ2fromu1} and \eqref{eqLinu2} together provide an implicit equation for $\psi_2$. 
With this result, (\ref{eqUpperLin}) and (\ref{eqLowerLin}) give the shapes of the upper and lower interfaces for any given $\psi_1$ in the limit of small interfacial slopes.
The shape of a liquid bridge determined using this method is shown in figure \ref{figTwoSolutions}(a) as the magenta and blue dashed curves. It is a very good approximation for the solution obtained using the nonlinear Laplace--Young equation.

\subsubsection{Approximation of the elliptic integrals}

\label{ellipticApprox}

The solution to the nonlinear Laplace--Young equation was given as a function of elliptic integrals in \eqref{eqfgboth}. Here we introduce an approximation to these integrals for the lower meniscus in order to obtain simpler relationships that can describe the meniscus shapes and the trapping behaviour. Since $g_2(v_2)=0$ according to \eqref{eqJBoth}, the relationship \eqref{eqfgboth} reduces for the lower meniscus to
\begin{align}
\label{eqfg}
f_2(y) = -\sgn(y)~g(y)+[\sgn(y)-\sgn(v_2)]~g(0).
\end{align}
We now use the values of $p$ and $q$ for the lower meniscus, \eqref{eqp} and \eqref{eqq}, on \eqref{eqJBoth} to obtain
\begin{align}
\label{eqJLo}
g_2&(y) =  |v_2| E \left[ \frac{1}{2} \cosinv \left( 1+ \frac{y^2-v_2^2}{2}\right), \left(\frac{2}{v_2}\right)^2 \right]  \nonumber\\
&-\left( |v_2|-\frac{2}{|v_2|} \right) F \left[ \frac{1}{2} \cosinv \left( 1+ \frac{y^2-v_2^2}{2}\right), \left(\frac{2}{v_2}\right)^2 \right].
\end{align}
The elliptic integrals in the above equation can be replaced using the following transformation formulae \citep{Byrd1971}:
\begin{align}
F(\sigma,k)&=\frac{1}{\sqrt{k}}F\left(\beta, \frac{1}{k}  \right), \\
E(\sigma,k)&=\sqrt{k}\left[ E\left(\beta, \frac{1}{k}  \right)- \left(1-\frac{1}{k} \right) F\left(\beta, \frac{1}{k}  \right) \right],
\end{align}
where
\begin{align}
\beta = \sininv( \sqrt{k} \sin \sigma).
\end{align}
This produces
\begin{align}
\label{eqG}
g_2(y) =  &2 E \left[  \sininv \sqrt{1- \frac{y^2}{v_2^2}}~, \left(\frac{v_2}{2}\right)^2 \right] \nonumber\\
&- F \left[  \sininv \sqrt{1- \frac{y^2}{v_2^2}}~, \left(\frac{v_2}{2}\right)^2 \right].
\end{align}
Substitution of $y=0$ gives
\begin{align}
\label{eqG0}
g_2(0) =  &2 E \left[ \left(\frac{v_2}{2}\right)^2 \right] - K \left[\left(\frac{v_2}{2}\right)^2 \right],
\end{align}
where $E(k)$ and $K(k)$ are complete elliptic integrals.
\citet{Byrd1971} gives series approximations for these functions. Using the first term of each series, we obtain
\begin{align}
\label{eqGs}
g(y) \approx  &2  \sqrt{1- \frac{y^2}{v_2^2}} - \ln \left[ \frac{|v_2|-\sqrt{v_2^2-y^2}}{|y|}\right]
\end{align}
and
\begin{align}
\label{eqG0s}
g(0) \approx \frac{\pi}{4}(2+\sqrt{4-v_2^2})-\frac{2 \pi}{2+\sqrt{4-v_2^2}}.
\end{align}
This expression for $g(0)$ is then used in the next section to determine an approximate solution for the shape of the lower meniscus.

\subsubsection{Solution for large liquid volumes, $|\psi_2| \to \pi/2$}
\label{solutionLargePsi2}

The solution given in section (\ref{solutionSmallPsi2}) is applicable for small $|\psi_2|$ and therefore represents liquid bridges that contain only a small liquid volume. We now introduce a solution for liquid bridges where $\psi_2$ is close to $\pi/2$, and where the trapped volume is large and hence,  to counterbalance the weight of the liquid, the vertical component of the surface tension force is high. In this regime, we focus on the largest liquid bridges, for which $v_2<0$ and $u_2>0$.

The shape of the upper part of the lower meniscus, near the contact points, may most readily be described by $x=h(y)$ with $h_y \ll 1$. The linearized  Laplace--Young equation for this regime is therefore
\be
y=h_{yy},
\ee
which we may solve to obtain
\begin{equation}
x = h(y)=\frac{1}{6}y^3+ c_1 y + c_2,
\label{eqLinLoUp2}
\end{equation}
where $c_1$ and $c_2$ are constants. These constants can now be constrained by our solutions to the nonlinear Laplace--Young equation. We first recall the constrains \eqref{eqa1} and \eqref{dfdy}, obtained in the solution of the nonlinear Laplace--Young equation, which gives
\begin{equation}
f_y(0) = \frac{1-\frac{1}{2}v_2^2}{|v_2|\sqrt{1-\frac{1}{4}v^2}}.
\label{dfdy0}
\end{equation}
We use the conditions  $h_y(0) = f_y(0)$ and  $h(0)=f(0)$, where $f(0)$ is given by the approximation \eqref{eqG0s}, to determine $c_1$ and $c_2$. Thus, we have
\begin{align}
h(y) = &\frac{1}{6}y^3 + \frac{1-\frac{1}{2}v_2^2}{|v_2|\sqrt{1-\frac{1}{4}v_2^2}}~ y + \frac{\pi}{4}(2+\sqrt{4-v_2^2})\nonumber\\
&-\frac{2 \pi}{2+\sqrt{4-v_2^2}}~,
\label{eqh}
\end{align}
which we may combine with the approximation for the upper meniscus determined for $|\psi_1| \rightarrow 0$ in section \ref{solutionSmallPsi1}. This approximation along with \eqref{u1u2psi} produces an expression for $u_2$, \eqref{equ2fromu1}. Combination of this expression with \eqref{equivi} gives
\begin{widetext}
\begin{align}
\label{eqv2omegaLin}
&v_2^2 = \Big\{ \sgn(\psi_1) 
\sqrt{\sin^2 \psi_1 ~\cosech^2 \left[d-R~ \sin (\theta-\psi_1)\right] + 2 \left(1- \cos \psi_1 \right) }
 - R \left[\cos (\theta-\psi_1) + \cos (\theta+\psi_2) \right] \Big\}^2 + 2 (1-\cos \psi_2).
\end{align}
\end{widetext}
We then use the boundary condition given in \eqref{bound3All}, that the fluid intersects the cylinder
\be
h(u_2) = d-R \sin (\theta+\psi_2),
\label{eqBoundary3Psi}
\ee
along with $u_2$ given by \eqref{equ2fromu1} and $v_2$ given by \eqref{eqv2omegaLin}, to get an equation which may
be solved to determine $\psi_2$. We note that  $h(y)$ is a good approximation for the upper part of the lower meniscus, as demonstrated in figures \ref{figShapes1}.

\begin{figure*}[tp]
\centering
\includegraphics [width=13cm] {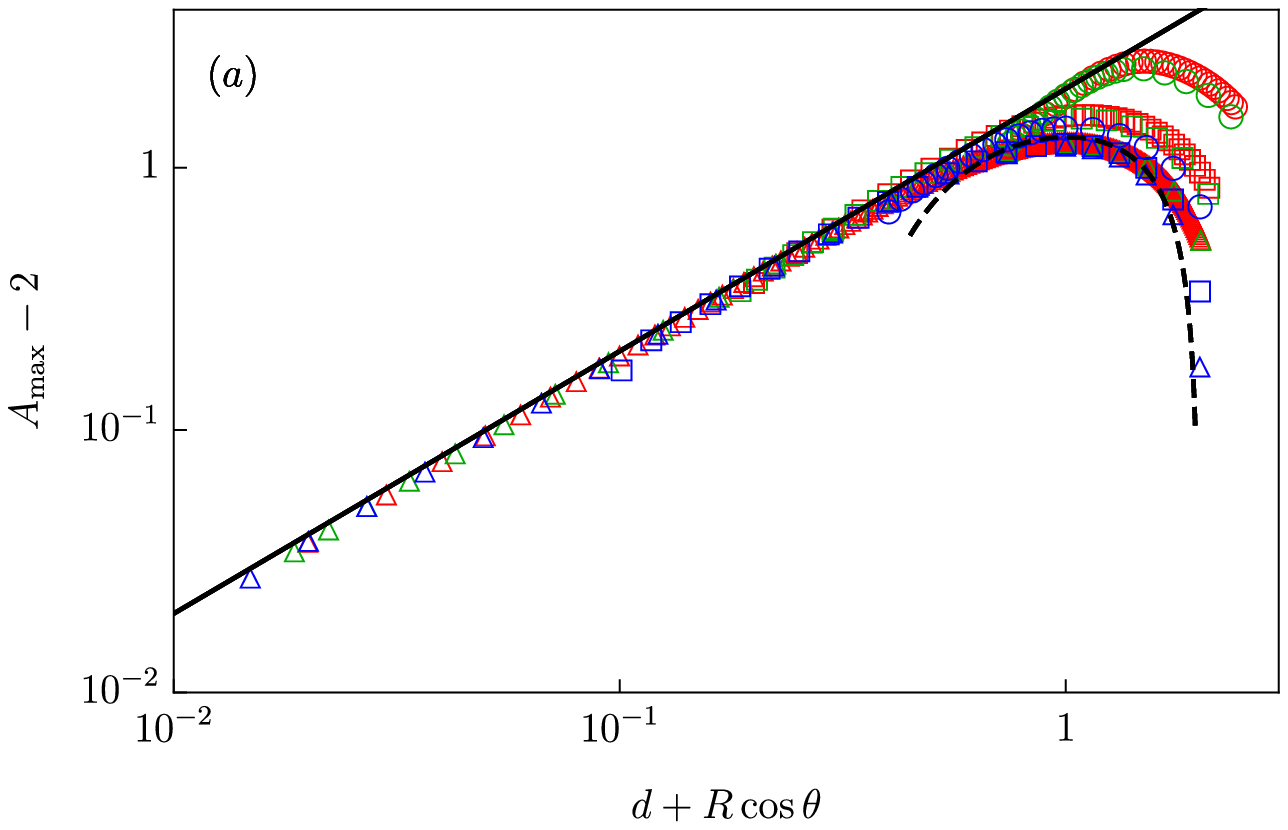}\\
\vspace{11 mm}
\includegraphics [width=13cm] {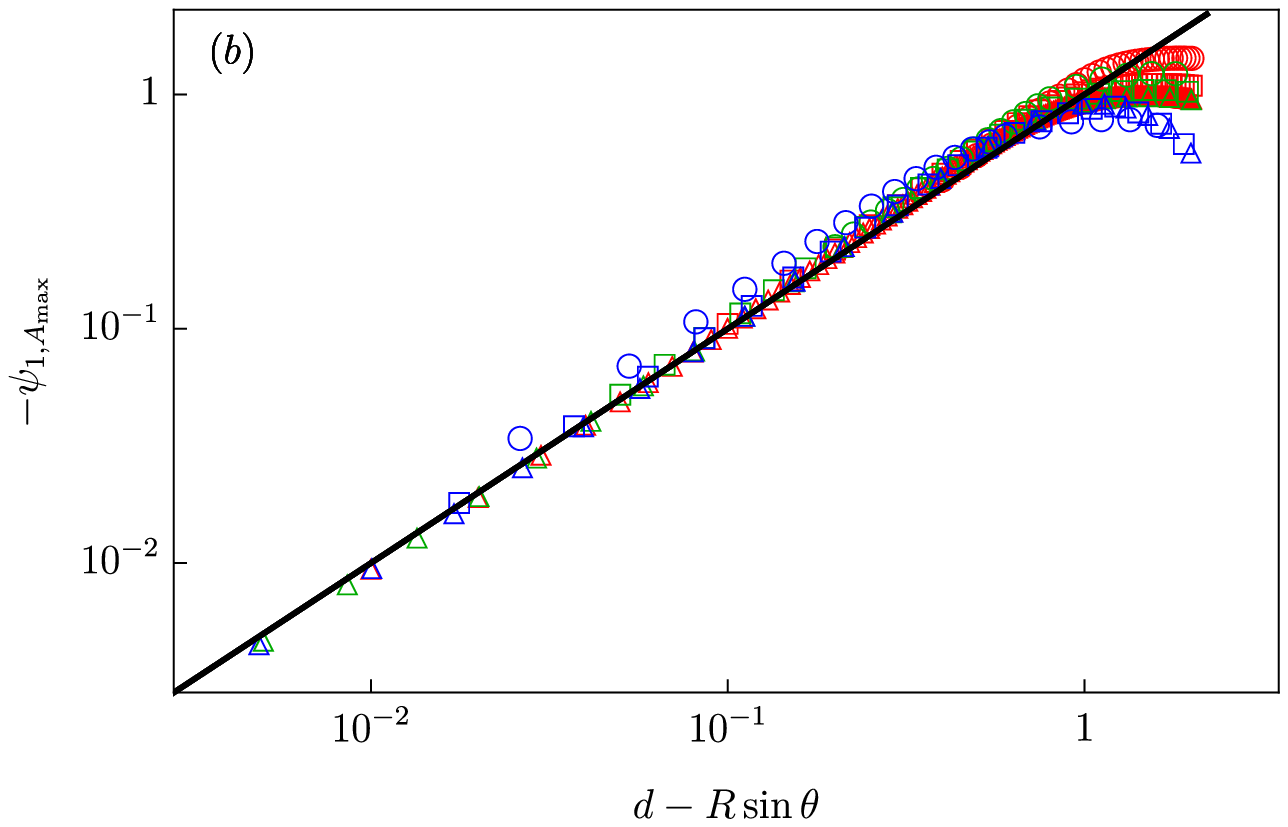}
\caption{
$(a)$ The maximum trapping capacity (the cross sectional area of the largest liquid bridge, $A_{\mathrm{max}}$) between pairs of horizontal cylinders. $(b)$ The value of $\psi_1$ for the largest liquid bridges. Symbols show results obtained 
for different values of $R$, $\theta$ and $d$ by numerically maximising $A$ \eqref{eq:bridgearea}, determined using the solution of the  nonlinear Laplace--Young equation, with respect to $\psi_1$.
Each marker represents a cylinder radius. Triangles ($\triangle$) denote $R=0.01$, 
squares ($\Box$) denote $R=0.1$ and 
circles ($\Circle$) denote $R=0.4$. 
Colours represent different contact angles. 
Red symbols (\textcolor{red}{$\triangle,\Box,\Circle$}) represent $\theta=0$, 
green symbols (\textcolor{myGreen}{$\triangle,\Box,\Circle$}) represent $\theta=\pi/6$ and
blue symbols (\textcolor{blue}{$\triangle,\Box,\Circle$}) represent $\theta=\pi/2$.
The black curves are approximations for the maximal trapping parameters. In $(a)$, the black solid line (--)  denotes \eqref{eqAmax}, which is valid for small $d$, and the black dashed curve (- -) denotes \eqref{eqAmaxLc}, which is valid when $d$ is close to 1. The black solid line in $(b)$ is the equation \eqref{eqW1c}. The approximate solutions describe the maximal trapping behaviour very well for small $R$ ~($R \ll 1$). Both  $A_{\mathrm{max}}$ and $\omega_{1,A_{\mathrm{max}}}$ are linearly related to $d$ when $d \ll 1$.
}
\label{figAmaximize}
\end{figure*}

Once $\psi_2$, and hence $v_2$, are determined, the shape of the lower part of the lower meniscus can be obtained approximately. The meniscus slopes in this regime are small relative to the $x$ axis, and therefore the linearized Laplace--Young equation, \eqref{linLYELow1}, is applicable. Solution with the boundary condition $k(0)=v_2$ gives $y=v_2 \cos x$, or
\be
x = \cosinv \left(y/v_2 \right).
\label{lowerMeniscusLowerPart}
\ee
We combine the solutions for the upper part of the lower meniscus \eqref{eqh} and lower part of the lower meniscus \eqref{lowerMeniscusLowerPart} to produce the following empirical expression for the meniscus shape,
\begin{align}
x=&\tanh \left[\frac{7}{4}(y-v_2) \right]h(y)\nonumber\\
&+ \left\{1- \tanh [2(y-v_2)]\right\}  \cosinv \left(\frac{y}{v_2}\right),
\label{compositeSol}
\end{align}
which is valid for the entirety of the lower meniscus as shown in figures \ref{figTwoSolutions} and \ref{figShapes1}.

\section{The maximal trapping capacity}

A quantity of significant interest in a variety of physical settings is the volume of fluid that may be trapped as a function of the imposed geometry and material properties through the apparent contact angle. Here we calculate the trapping capacity, which in our two--dimensional geometry is equivalent to the cross--sectional area. We then determine the maximum achievable trapping capacity at a given separation between the cylinders and the separation at which the largest liquid bridge can be produced. 

\begin{figure*}[t!]
\centering
\includegraphics [width=10cm] {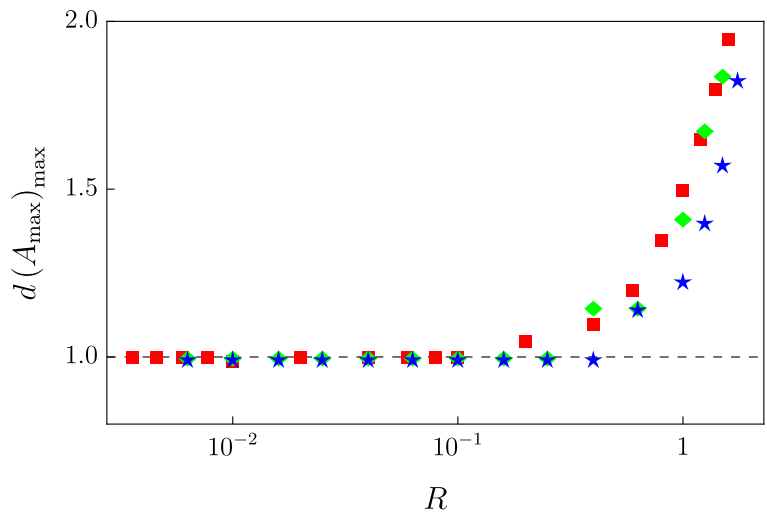}
\caption{For small cylinders, the maximum trapping capacity $A_{\mathrm{max}}$ maximizes at $d \approx 1$. Figure shows the value of $d$ at which the maximum trapping capacity occurs, obtained using the nonlinear Laplace--Young equation. Red squares ($\textcolor{red}{\blacksquare}$) are for $\theta = 0$, green diamonds ($\textcolor{green}{\blacklozenge}$) are for $\theta = \pi/6$ and blue stars  ($\textcolor{blue}{\bigstar}$) are for $\theta = \pi/2$.
}
\label{figAmax}
\end{figure*}

\subsection{The maximum trapping capacity at a given separation}

The cross sectional area $A$ of a liquid bridge can be determined using a force balance considering the liquid weight and the forces of surface tension and hydrostatic pressure. 
\begin{align}
  \label{eq:bridgearea}
  A = &-2(\sin\psi_1+\sin\psi_2) \nonumber\\
&+ 2R~u_1\left[ \sin(\theta-\psi_1)-\sin(\theta+\psi_2) \right] \nonumber\\
&+ R^2\Bigg[\psi_2-\psi_1+2\theta-\pi + 2\cos(\theta-\psi_1)\sin(\theta+\psi_2) \nonumber\\
&~~~~~~~~~~- \frac{\sin 2(\theta-\psi_1)-\sin 2(\theta+\psi_2)}{2}  \Bigg].
\end{align}
The quantities $\psi_2$ and $u_1$ in this equation can be determined as functions of $\psi_1$ using the solution of the nonlinear Laplace--Young equation described in section~\ref{sec:exactsolution}. By numerical maximisation of $A$ with respect to $\psi_1$, the  maximum trapping capacity $\left(A_{\mathrm{max}} \right)$ and $\psi_1$ that produces this trapping capacity $\left(\psi_{1,\mathrm{A_{max}}} \right)$ can be determined for a given combination of $R,~ \theta$ and $d$. Two representative liquid bridges, corresponding to $A_{\mathrm{max}}$ for different values of $d$, are shown in figures \ref{figShapes1}$(c)$ and $(d)$. This solution process was repeated for a range of $R,~\theta$ and $d$, and the behaviour of $A_{\mathrm{max}}$ and $\psi_{1,\mathrm{A_{max}}}$ were analysed. The results are shown by symbols in figure \ref{figAmaximize}$(a)$ and $(b)$. 

Figure \ref{figAmaximize}$(a)$ shows that the maximal trapping capacity, $A_{\mathrm{max}}$ is linearly proportional to the separation, $d$, when $R \ll 1$ and $d \ll 1$. This relationship can be explained using the approximate solution derived in section (\ref{sectionApprox}). We define $2s_i$ as the distance between the contact points of a meniscus, that is,
\be
s_i = d-R \sin \omega_i,
\label{si}
\ee
so that when $d,~R \ll 1$, $s_i \ll 1$ for all $\omega_i$.
In this regime, the shape of the upper meniscus has a nearly constant radius of curvature $-s/\sin \psi_1$.
As a result we have
\be
|u_1-v_1| \le s_1.
\ee
Since $s_1 \ll 1$ the amount of liquid trapped above the $y=v_1$ is negligible and since $R \ll 1$ almost all the liquid is trapped as a droplet hanging below $y=u_2$. 
The cross--sectional area of the part of the liquid bridge below $y=u_2$ is determined by balancing the non--dimensionalized weight of the liquid $A$ with the force of surface tension given by $-2 \sin \psi_2$ and the force of hydrostatic pressure given by $2 u_2 s_2$,
\begin{equation}
A = 2\left( -\sin \psi_2+ u_2 ~s_2  \right).
\label{eqA}
\end{equation}
The surface tension force acting on the liquid bridge is more significant compared to the force of hydrostatic pressure since $s_2 \ll 1$. $A$ is therefore maximized when
\be
\psi_2  \approx -\pi/2, 
\label{psi2is90}
\ee
which is the meniscus slope angle that maximizes the vertical component of the force of surface tension.
Using \eqref{psi2is90} and $d-R \sin (\theta+\psi_2) =s_2$ on \eqref{eqBoundary3Psi} and replacing $q_2$ and $v_2$ using \eqref{eqq} and \eqref{equivi} respectively, we obtain an equation for $u_2$
\begin{align}
\left( \frac{1}{6}-\frac{1}{\sqrt{4-u_2^4}} \right)u_2^3 & \nonumber\\
+ \frac{\pi}{4}\left(2+\sqrt{2-u_2^2}\right) & \nonumber\\
-\frac{2\pi}{2+\sqrt{2-u_2^2}}&=s_2.
\label{equ2Amax}
\end{align}
As $s \rightarrow 0$, 
\begin{equation}
u_2 \approx 1
\label{u2is1}
\end{equation}
is an approximate solution for \eqref{equ2Amax}.
Substitution of \eqref{psi2is90} into \eqref{psi2} gives $\sin \omega_2 = -\cos \theta$, which in combination with \eqref{si} produces
\begin{equation}
s_2 =  d+R \cos \theta.
\label{sinOmegaIsCosTheta}
\end{equation}
Substitution of \eqref{psi2is90}, \eqref{sinOmegaIsCosTheta} and \eqref{u2is1} into \eqref{eqA} produces
\begin{equation}
A_{\mathrm{max}} \approx 2 (1+ d + R \cos \theta).
\label{eqAmax}
\end{equation}
This is plotted by the black line shown in figure \ref{figAmaximize}$(a)$. It is a good approximation for small cylinders at close range. 

We also observe a linear relationship between  and $\psi_{1,A_\mathrm{max}}$ for small $R$ and $d$ in figure \ref{figAmaximize}$(b)$. This relationship can also be verified using the approximate solutions to the Laplace--Young equation. 
If $|\psi_1|$ is small, \eqref{eqUpperLin} is valid throughout the upper meniscus, which gives
\begin{equation}
u_1 = -\sin \psi_1 ~\coth (d-R \sin \omega_1).
\label{equ1}
\end{equation}
Since $R \ll 1$ we have 
\be
u_1 \approx u_2,
\label{u1equ2}
\ee
which gives $u_1 \approx 1$ due to \eqref{u2is1}. Using this result on \eqref{equ1}, we obtain
\begin{equation}
-\sin \psi_{1,A_\mathrm{max}} = \tanh (d-R~ \sin \omega_{1,A_\mathrm{max}}).
\end{equation}
Since $d-R ~\sin \omega_{1,A_\mathrm{max}} =s_1 \ll 1$, the above equation gives
\begin{equation}
-\psi_{1,A_\mathrm{max}} \approx d-R~ \sin \omega_{1,A_\mathrm{max}},
\label{eqW1b}
\end{equation}
where $\psi_1$ and $\omega_1$ are related by \eqref{psi1},
which gives 
\begin{equation}
\omega_1 \approx \theta
\label{eqW1d}
\end{equation}
for small $\psi_1$.
Substitution of \eqref{eqW1d} to \eqref{eqW1b} gives the relationship
\begin{equation}
 -\psi_{1,A_\mathrm{max}} \approx d-R \sin \theta,
\label{eqW1c}
\end{equation}
which is plotted by the black line in figure \ref{figAmaximize}$(b)$.
This result approximates the exact solution very well when the cylinder radius and inter--cylinder radius are small compared to the capillary length.

\subsection{The separation which maximizes the trapping capacity}

\begin{figure*}[t!]
\centering
\includegraphics [width=8cm] {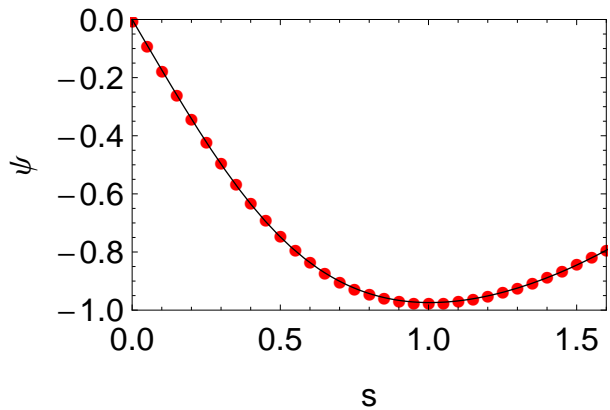}
\caption{
The expression for $\psi$ given in \eqref{eqPsiFromS} (solid curve: --) is compared with $\psi$ obtained using a numerical solution of \eqref{eqmpsi} (red symbols: $\textcolor{red}{\bullet}$). The figure shows that the analytical expression is an accurate solution for  \eqref{eqmpsi}.
}
\label{figPsiVsS}
\end{figure*}

\begin{figure*}[t!]
\centering
\includegraphics [width=6.5cm] {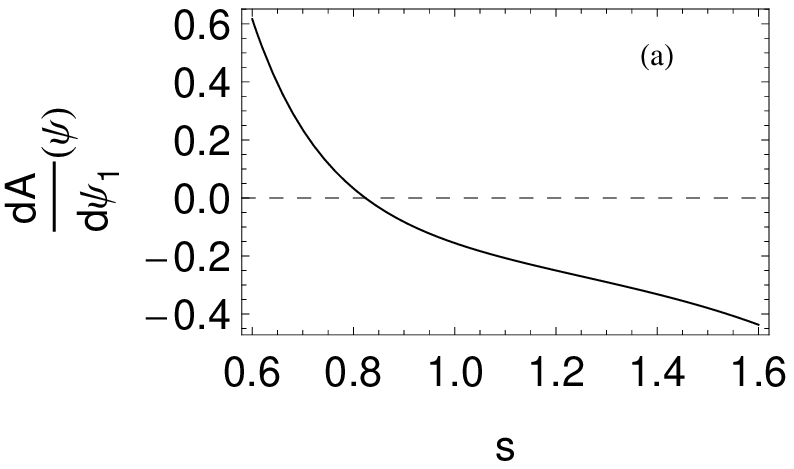}
\includegraphics [width=6.5cm] {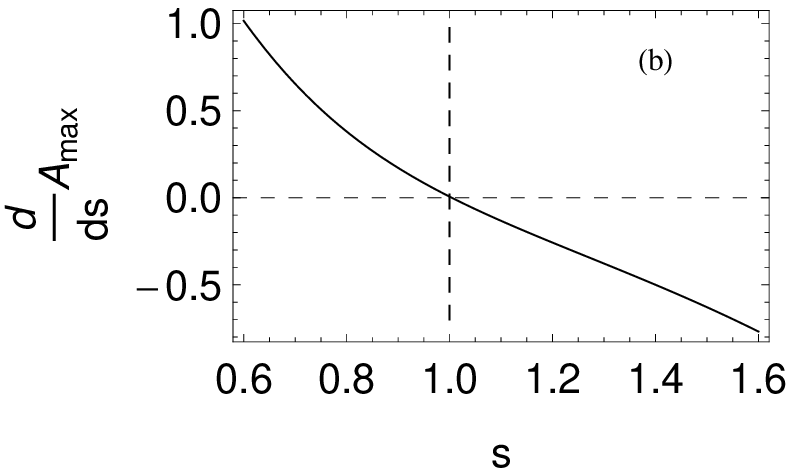}
\caption{(a) shows $\upd A / \upd \psi_1$ calculated using \eqref{eqDADpsi} at $\psi_1 = \psi_2 = \psi$. The derivative is negative around $s=1$ and beyond. Since $u_1 \approx u_2$ and the two menisci should not intersect, the minimum possible value of $\psi_1$ is $\psi$. The negative derivative means that $A$ maximizes when $\psi_1 = \psi_2 =\psi$. (b) shows the derivative of $\upd V / \upd s$ calculated at $\psi_1 = \psi_2 =\psi$. $A$ maximizes at $s=1$.
}
\label{figdA}
\end{figure*}

Figure \ref{figAmaximize}$(a)$ shows that the maximum trapping capacity $A_{\mathrm{max}}$ as a function of $d$ is increasing when $d \ll 1$ and decreasing for large $d$. Figure \ref{figAmax} plots the value of $d$ in which $A_{\mathrm{max}}$ reaches a maximum $\left( ~d~{\left(A_{\mathrm{max}}\right)_{\mathrm{max}}} ~\right)$ as a function of $R$ for different values of $\theta$. Interestingly, it shows that $d~{\left(A_{\mathrm{max}}\right)_{\mathrm{max}}} = 1$ when $R \ll 1$ for all $\theta$. In this section, we analytically explain this result based on the approximate solutions obtained earlier for the liquid bridge geometry.

We assume that \eqref{eqUpperLin} gives a sufficiently good approximation for $u_1$
\be
u_1 = -\sin \psi_1 \coth s,
\label{u1linLYEsmallR}
\ee
where $s = s_i \approx d$ which is valid when $R \rightarrow 0$ according to \eqref{si}. 
For small $R$, we also have $u_2 \approx u_1$ which gives
\begin{equation}
u_2 \approx -\sin \psi_1 \coth s,
\end{equation}
and \eqref{equivi} then gives
\be
v_2^2 \approx \coth ^2s ~\sin ^2\psi_1 -2 \cos \psi_2+2.
\ee
Substitution of the above two expressions obtained for $u_2$ and $v_2$ into \eqref{eqBoundary3Psi} yields
\begin{equation}
m(\psi_1,\psi_2,s)=0,
\label{eqmpsi}
\end{equation}
where
\begin{align}
\label{eqM}
m(\psi_1,\psi_2,s)=&-\frac{1}{6} \coth ^3s \sin ^3 \psi_1 \nonumber\\
&-\frac{\coth s \sin \psi_1 \left(2 \cos \psi_2 - \coth ^2 s~ \sin ^2 \psi_1 \right)}{\sqrt{4-\left(\coth ^2 s \sin^2 \psi_1 -2 \cos \psi_2\right)^2}} \nonumber\\
&+\frac{1}{4} \pi  \left(2+\sqrt{2 \cos \psi_2-\coth ^2s \sin ^2 \psi_1+2}\right) \nonumber\\
&-\frac{2 \pi}{2+\sqrt{2 \cos \psi_2-\coth ^2s \sin ^2\psi_1+2}}\nonumber\\
&-s.
\end{align}

Since the contact points of the upper and lower menisci are very close to each other ($u_1 \approx u_2$), we need $\psi_1 \ge \psi_2$ to avoid the two menisci intersecting each other. 
We now consider the limit $\psi_1 = \psi_2 = \psi$, where \eqref{eqmpsi} is written as 
\begin{equation}
m(\psi,s)=0.
\label{eqmpsiEq}
\end{equation}

To solve for $\psi$, $m(\psi,s)$ is expanded in a first order power series
\begin{equation}
m(\psi,s)= m(\psi_0,s)+(\psi-\psi_0)~m_\psi(\psi_0,s).
\label{eqExpansionm}
\end{equation}
A numerical solution of \eqref{eqmpsiEq} shows that $\psi \approx -\frac{7}{4} s$ as $s \rightarrow 0$ and  $\psi \approx -1$ as $s \rightarrow 1$.  We therefore select $\psi_0$ in \eqref{eqExpansionm} as
\begin{align}
\psi_0 (s)&= -\frac{7}{4}s(1-s)-s^2\\\nonumber
&=\frac{1}{4}\left(3s^2-7s\right),
\end{align}
which gives the solution
\begin{equation}
\psi(s) = \frac{1}{4}\left(3s^2-7s\right) -
\frac{m\left[\frac{1}{4}\left(3s^2-7s\right),s\right]}{m_\psi\left[\frac{1}{4}\left(3s^2-7s\right),s\right]}.
\label{eqPsiFromS}
\end{equation}

To test the accuracy of the solution for $\psi$ given by  \eqref{eqPsiFromS}, it is compared with the numerical solution of  \eqref{eqmpsiEq}. As shown in figure \ref{figPsiVsS}, the accuracy of the analytical approximation is very good for a wide range of $s$. 

The force of hydrostatic pressure exerted by small cylinders on a liquid bridge is negligible compared to the force of surface tension because the solid--liquid contact area is small. The cross--sectional area of the liquid bridge can therefore be calculated by balancing the surface tension force with the weight 
\be
A = -2 \left( \sin \psi_1 +\sin \psi_2 \right).
\ee
When $\psi_1=\psi_2=\psi$ for a given $s$ we have
\begin{equation}
\frac{\upd A}{\upd \psi_1} [\psi(s)] = -2\cos \psi \left( 1 + \frac{\partial \psi_2}{\partial \psi_1}[\psi(s),s] \right),
\label{eqDADpsi}
\end{equation}
where
$ \partial \psi_2 / \partial \psi_1$ is obtained as a function of $\psi_1 , \psi_2$ and $s$ by differentiating \eqref{eqmpsi} with respect to $\psi_1$.

Figure \ref{figdA} (a) shows that $ \partial \psi_2 / \partial \psi_1 [\psi(s)]$ is negative around $s=1$, which means the trapping capacity for a given separation of around 1 is maximized when $\psi_1=\psi_2=\psi(s)$. The maximum trapping capacity is therefore given by 
\be
A_{\mathrm{max}} = -4 \sin \psi(s).
\label{eqAmaxLc}
\ee
For small cylinders, \eqref{eqAmaxLc} gives the value of $A_{\mathrm{max}}$  at far range while \eqref{eqAmax} explains the behaviour at short range as shown in figure \ref{figAmaximize}$(a)$.

Differentiation of \eqref{eqAmaxLc} gives
\be
\frac{\upd A_{\mathrm{max}}}{\upd s} = -4 \cos{\psi} \frac{\upd \psi}{\upd s},
\ee
\noindent while $\psi$ and $\upd \psi / \upd s$ can be obtained from \eqref{eqPsiFromS}. According to figure \ref{figdA} (b), $A_{\mathrm{max}}$ is a maximum when $s=1$. 
According to the results in figure \ref{figAmax}, which are obtained by solving the nonlinear Laplace--Young equation, $A_{\mathrm{max}}$ maximizes at $d \approx 1$ when $R \ll 1$. Both these results are similar since $d \approx s$ for small $R$.
 
\section{Conclusions}

We present exact solutions to the nonlinear Laplace--Young equation to determine the equilibrium shape of a liquid bridge trapped between a pair of infinitely long horizontal cylinders. We also introduce several simpler solutions that approximate the exact solutions very well. 

Both the exact and approximate solutions show that the maximum amount of liquid that can be trapped in a given system and the conditions of this maximisation can be approximated by a few simple relationships when the cylinder radius is small compared to the capillary length ($\lc$).  Regardless of the contact angle, the largest liquid bridges form when the inter--cylinder distance is approximately $2~ \lc$. 
If the inter--cylinder distance is small compared to $\lc$, the maximum amount of liquid held by a pair of cylinders is given by the equation $a_{\mathrm{max}} \approx 2\lc~ (1+D+r~\cos \theta)$, in which $a$ is the cross--sectional area of the liquid bridge, $2 D$ is the inter--cylinder distance, $r$ is the cylinder radius and $\theta$ is the contact angle. At this maximum trapping, the meniscus slope angle of the upper interface of the liquid bridge can be approximated by the linear relationship $\psi_{1,a_{\mathrm{max}}} \approx \left(r~\sin \theta - D\right)/\lc$.

The solutions we present here can be extended to determine the equilibrium of fluid ganglia or stringers trapped in a solid matrix, enclosed by a different non--mixing fluid. Although such systems have been studied neglecting gravitational effects \citep{Niven2006}, an analysis considering the weight of the fluid can help determine the residual trapping capacity of a porous medium. It can also be used to characterise deformations of the solid support induced by the surface tension forces from fluid ganglia and any fluid movement that result from this. This is a significant factor in trapping by a flexible solid support, as shown by \citet{Duprat2012} for the case of small liquid bridges between cylinders.

\section*{Acknowledgment}

We thank Raphael Blumenfeld for many useful discussions. This work is funded under the EU TRUST consortium. HEH was partially funded by a Leverhulme Emeritus Professorship during this research. JAN is partially supported by a Royal Society University Research Fellowship.


\bibliographystyle{unsrtnatH}

\bibliography{library3}
\end{document}